# An Empirical Investigation of Correlation between Code Complexity and Bugs

Changqi Chen
Email: changqc@uci.edu

*Abstract*—There have been many studies conducted on predicting bugs. These studies show that code complexity, such as cyclomatic complexity, correlates with the presence of bugs in code. In this paper, we intend to find the correlation between path complexity and bugs. We found that 1) For simple bugs, there is no strong correlation between the path complexity and the presence of bugs; 2) For complex real-world bugs, though not strong, path complexity has a higher correlation with the presence of bugs than cyclomatic complexity and NPATH complexity. These results are useful for researchers to use the path complexity for building bug prediction models. Moreover, path complexity can be used as a guiding mechanism for test generation.

## I. INTRODUCTION

Bug prediction is one of the most critical research areas in software engineering. Researchers have tried to identify and validate metrics that have a high correlation with the presence of bugs in code. These metrics are helpful as bug predictor in many testing techniques such as fault localization [1,2] and Feedback-directed Random Testing (FRT) [3].

Many of the existing techniques concentrate on the number of execution paths instead of other heuristics such as branch coverage or statement coverage [4]. There are different complexities compute the execution paths, such as cyclomatic complexity [5] and path complexity [4].

In this paper, we try to find the correlation between the path complexity and the presence of bugs in code. We are interested in path complexity because it is different in its representation. It, rather than provides a constant number, provides the number of execution paths concerning execution depth [4]. We compare the path complexity with cyclomatic complexity [5] and NPATH complexity [6], which both present the complexity as a constant number.

We answer to research questions about the correlation between path complexity and the presence of bugs based on two different kinds of bug dataset: 1) 395 simply fixed bugs in Defects 4j 2) 35 complex and more realistic bugs in BugSwarm. The specific research questions we answer in this paper are:

*RQ 1: Is there a correlation between the path complexity and bugs in Defects4j?*

*RQ 2: Is there a correlation between path complexity and more complex and realistic bugs?*

The remainder of this paper is organized as follows: Section II presents the related work. Section III presents our methodology. Section IV answers to the research questions with results and analysis. Section V discusses the threats to validity, and Section VI presents the conclusions.

## II. RELATED WORKS

### A. Path complexity

Bang et al. in their recent work, presented the path complexity [4]. The path complexity represents the complexity as a symbolic expression on a single variable n which denotes the execution depth [4]. In their paper, they showed the method for computing the path complexity, compared the path complexity with other complexities and provided the tool PAC to compute the path complexity automatically. The results of his experiment showed that cyclomatic complexity and NPATH complexity could not differentiate between methods with the constant number of execution paths and methods with iterations or recursions. The reason is that those complexities are always represented as constant numbers. On the other hand, since path complexity can be represented as expressions and grow exponentially with the number of execution paths in method, it can be a better choice to access the difficulty of achieving path coverage [4].

### B. Bug prediction approaches

There have been different kinds of approaches in bug prediction. For example, Nagappan and Ball proposed to use relative code churn (the amount of change to the system) as a predictor of bugs [13]. Hassan introduced the entropy of changes, a measure of the complexity of code changes, as one predictor [14]. These approaches require both the recently changed and the current files to predict the bugs. Another kind of approaches, including our approach, analyzes the current state file in more detail and predict the bugs. Ohlsson et al. used several graph metrics, including cyclomatic complexity as the predictor [15].

## III. METHODOLOGY

To find the correlation between the path complexity and the presence of the bugs, we define the following research questions.

*RQ 1: Is there a correlation between the path complexity and bugs in Defects4j?*

We first use the 395 bugs in Defects4j [7] as the dataset and compute their cyclomatic complexity, NPATH complexity and path complexity for both the buggy version and fixed version. If the path complexity has a stronger correlation with the presence of bugs than the other two, then we can conclude that path complexity can be a good guiding mechanism for test generation.

Defects4j is a bug dataset that contains 395 bugs in six open-source Java projects: Commons Lang, Commons Math, Closure Compiler, Joda-Time, Mockito, and JfreeChart. The bugs in Defects 4j have three essential characteristics: 1) related to source code 2) reproducible 3) isolated [9].

In this paper, we compute three different complexities, cyclomatic complexity, NPATH complexity and path complexity, for each bug and compare their correlation with both the buggy version and fixed version. Cyclomatic complexity computes the maximum number of linearly independent paths in the CFG from the entry node to the exit node [11]. It does not consider the execution depth for loop or recursion. NPATH complexity simplifies this problem by always counting the paths that execute the loop or recursion once or zero time [12]. Cyclomatic complexity and NPATH complexity both represent the complexities as constant numbers. They either ignore the problem of execution depth or simplify it. Path complexity on the other hand, represents complexity as a symbolic expression on a single variable n, denoting the execution depth [4]. We compute the cyclomatic complexity and NPATH complexity using Understand (SciTools) [17] and compute the path complexity using PAC [4].

*RQ 2: Is there a correlation between path complexity and more complex and realistic bugs?*

We identify the largest and realistic bug dataset BugSwarm [8]. It contains more than 3,000 reproducible bugs in projects written in either Java or Python. We need the bugs to be not only complex enough but also isolated and reproducible. Most bugs in BugSwarm are not isolated. So, we filtered the 789 bugs in BugSwarm whose language is Java and patch size is greater than 20. We manually make sure the bugs we selected have all changed lines in a single method and patches do not include unrelated changes. The patch size is the sum of added, removed and modified lines in the patch. We end up having 35 qualified bugs. We used them as the bugs database to reproduce the analysis in RQ1.

We use Kendall's rank correlation tau value to evaluate the correlation between the complexity and the presence of bugs. It measures the strength of the monotonic association between two variables [16]. The range of value is between 0 and 1 where 0 means no relationship and 1 means perfect relationship.

All the bugs data collected from dataset Defects 4j and BugSwarm, consisting of all complexities and the information about the bugs, is publicly available in an open-science repository[1].

## IV. RESULTS

In this section, we present the results of our two research questions:

*RQ 1: Is there a correlation between the path complexity and bugs in Defects4j?*

We compare three complexities' (cyclomatic complexity, NPATH complexity, and path complexity) Kendall's rank correlation tau values to evaluate the strength of association. For path complexity, since we cannot use a symbolic expression to compute Kendall value, we only use the value of the highest term in complexity. We initially use all the bugs in Defects4j as our database. We expected that more complex codes will have more bugs. However, we find that all three complexities have low Kendall values. Since path complexity cares about iterations and recursions in method, our intuition is that it should have a higher correlation with bugs in code. But in our results, it does not show any relatively higher association with bugs. Cyclomatic complexity has the highest Kendall value of 0.064. Table 1 contains the detailed results of the Kendall value of each of the complexities. This result demonstrates that there is no correlation between complexity and the presence of bugs in code. The results do not match our expectations so we decide to investigate the bugs in Defects4j.

We find research on the Defects4j dataset shows that the bugs in Defects4j are not complex enough. Since Defects4j was initially used for the mutation test, most of the bugs in it are very simple. 25% of bugs has at most two changed lines and 95% of bugs have at most 22 changed lines [9]. The maximum changed lines are only 54 lines [9]. Moreover, bugs are not realistic. For instance, 4 out of 6 Defects 4j projects are library rather than real-world Java software [10]. This result indicates that bugs in Defects 4j are simple so that the difference between the buggy version and the fixed version is small and the difference in complexity is negligible. We then choose only the bugs fixed by adding conditional branches from Defects4j and we end up having 191 bugs. The results show that all three complexities' Kendall values increase, and the path complexity has the highest Kendall value among three complexities.

---

[1] https://github.com/changqc7/SURP

Therefore, we decided to find another more complex and realistic bug database for our research.

Table 1 Correlation between code complexity and bugs in Defects4j (* indicates statistical significance)

|  | Cyclomatic complexity | NPATH complexity | Path complexity |
|---|---|---|---|
| Defects4j bugs | 0.064 * | 0.063* | 0.051 |
| Bugs fixed by adding conditional branches | 0.126 * | 0.122 * | 0.147 * |

*RQ 2: Is there a correlation between path complexity and more complex and realistic bugs?*

As mentioned in the Methodology section, we identified 35 bugs out of the 3000 bugs collected from the BugSwarm dataset. The Kendall values results demonstrate that all three complexities do better in predicting bugs in more realistic and complex codes. The path complexity has the highest Kendall value among three complexities. Table 2 contains the results for BugSwarm. This result indicates that path complexity can be a better bug predictor. However, the p-value, which determines the significance of the results, increases a lot in RQ2, which may cause by the limited sample size.

Table 2 Correlation between code complexity and bugs in BugSwarm (* indicates statistical significance)

|  | Cyclomatic complexity | NPATH complexity | Path complexity |
|---|---|---|---|
| Bugs in BugSwarm (35 bugs) | 0.080 | 0.094 | 0.106 |

## V. THREATS TO VALIDITY

**Bias due to sampling:** Our bugs sample has been from a single language – Java. This can be a source of bias, and our findings can be limited to programs in Java and not generalizable to programs in other languages.

**Bias due to tools used:** The complexity computation tool we used is PAC which implemented by Bang et al. Software bugs are common, and it cannot be ruled out in the analysis we performed so the accuracy of our results is relying on the PAC tool. However, the threat is minimal because 1) we use another tool Understand also to compute the cyclomatic complexity and compare the results with cyclomatic complexity from PAC. 2) we rely on the fact that the Replication Packages Evaluation Committee successfully evaluates the PAC tool and finds it meets expectations.

## VI. CONCLUSIONS

Prior research, which introduced the path complexity, only validated that path complexity is better in assessing the difficulty of achieving path coverage than other graph metrics in their research. Our contribution in this paper is that we identify the path complexity can be a better bug predictor for test generation. We also find that the bugs in dataset Defects 4j are simple. So, it is not an appropriate bug database for researchers to find the correlations between code complexity and bugs.